\documentclass[prc,twocolumn,showpacs,amsmath,amssymb,floatfix]{revtex4}
\usepackage{graphicx}
\usepackage{bm}
\usepackage{booktabs}
\usepackage{subfigure}
\usepackage[usenames]{color}

\newcommand{\beq}{\begin{equation}}
\newcommand{\eeq}{\end{equation}}
\newcommand{\be}{\begin{eqnarray}}
\newcommand{\ee}{\end{eqnarray}}

\begin{document}
\title{Weak response of neutron matter at low momentum transfer}
\author{Omar Benhar$^{1,2}$}
\author{Andrea Cipollone$^{1,2}$}
\author{Andrea Loreti$^{2}$}
\affiliation
{
$^1$ INFN, Sezione di Roma. I-00185 Roma, Italy\\
$^2$ Dipartimento di Fisica, ``Sapienza''  Universit\`a di Roma. I-00185 Roma, Italy \\
}
\date{\today}
\begin{abstract}
The Landau parameters obtained from the matrix elements of an effective interaction recently derived within 
the formalism of correlated basis functions have been used to carry out a study of the weak response of
neutron matter in the region of low momentum transfer. The proposed approach allows for a consistent 
description of different interaction effects and can be extended to describe matter at non vanishing temperature. 
The results show that interactions lead to a sizable enhancement of the neutrino mean free path in cold
neutron matter. The dependence of the mean free path on temperature and neutrino energy is also analyzed. 
\end{abstract}
\pacs{24.10.Cn,21.65.-f,25.30.Pt,26.60.-c}
\maketitle

\section {Introduction}
\label{intro}
The understanding of neutrino interactions with dense nuclear matter is critical to the description of a number of astrophysical processes. 
The results of many-body calculations carried out using the formalism of Correlated Basis Functions (CBF) \cite{shannon_2,shannon0,shannon,Benhar_Farina} have clearly 
shown that nucleon-nucleon (NN) correlations strongly affect the weak response of nuclear matter. Short-range correlations bring about a sizable quenching of the relevant transition matrix 
elements, with respect to the predictions of the Fermi gas model, while long-range correlations lead to the excitation of collective modes that turn out
to be dominant at low momentum transfer. In Refs. \cite{shannon,Benhar_Farina} short- and long-range correlations have been treated in a consistent 
fashion, using correlated wave functions and the cluster expansion technique to derive effective interactions from realistic nuclear hamiltonians. 
The approach based on these effective interactions allows for
a unified treatment of a variety of properties of nuclear matter relevant to astrophysical applications, such as, for example, the transport coefficients 
and the neutrino emission rates \cite{BV,BPVV,BFFV}.

An alternative approach to the study of neutrino interactions in nuclear matter is founded on the theory of normal Fermi liquid, in which the dynamics 
is described by a set of Landau parameters \cite{L1,L2,IW}. This formalism, while being strictly applicable only in the regime where the 
excitations of the system can be described in terms of {\em quasiparticles}, allows for a consistent treatment of collective excitations and can be
easily extended to non vanishing temperatures.

In this paper we use the Landau parameters obtained from the matrix elements
of the effective interaction of Ref. \cite{BV} to carry out a study of the weak response of pure neutron matter in the region of low momentum 
transfer.

Section \ref{veff} is devoted to a short description of the many-body approach employed to derive the effective interaction, while in Section \ref{landau_par}
we outline the main elements of Landau theory, and report the results of our calculation of the Landau parameters. 
In Section \ref{static} the internal consistency of our approach is tested by comparing the values of spin susceptibility and compressibility 
computed using Landau parameters to those obtained from the matrix elements of the CBF effective interaction. The density and spin-density
structure functions of cold neutron matter are discussed in Section \ref{energy}, while in Section \ref{mfp} we analyze the
neutrino mean free path and report the results of calculations carried out over a broad density range at temperatures up to $10$, assuming non degenerate 
neutrinos.
Finally, in Section \ref{concl} we summarize our findings and state the conclusions. 
  

\section{Formalism}
\label{forma}

\subsection{CBF effective interaction}
\label{veff}

The formalism of nuclear many-body theory provides a consistent framework, suitable for 
treating the non perturbative nature of NN interaction. 
Within this approach, nuclear matter is modeled as a collection of $A$ point-like particles, the dynamics of which 
are dictated by the hamiltonian
\beq
H=\sum_{i}\frac{{\bf p}_i}{2m }+\sum_{j>i}v_{ij}+\ldots \ ,
\eeq	 
where ${\bf p}_i$ and $m$ denote the momentum of the $i$-th nucleon and its mass, respectively, $v_{ij}$ is the 
NN interaction potential and the ellipses refer to the presence of interactions involving three or more nucleons.

The NN potential $v_{ij}$
reduces to the Yukawa one-pion exchange potential at large
distances, while its behavior at short and intermediate range is determined by
a fit of deuteron properties and NN scattering phase shifts.
 
Performing perturbative calculations in the basis of eigenstates of the non interacting system requires the replacement of the  \emph{bare} NN potential with a well behaved effective 
interaction. This is the foundation of the approach developed by Br\"{u}ckner, Bethe, and Goldstone (for a review, see, e.g., Refs. \cite{BBG,Baldo}),  in which $v_{ij}$ is replaced 
by the reaction matrix $G$, obtained by summing up the series of ladder diagrams describing NN scattering in the nuclear medium. 
The same philosophy has been followed in more recent studies, such as those carried out within the self-consistent Green's function 
approach \cite{Pol}.
 
Alternately, non perturbative effects can be taken into account replacing the states of the non interacting system, i.e. Fermi gas states $| n_{FG} \rangle$ in the case of 
 uniform nuclear matter,  with a set of {\em correlated states}, defined as (see, e.g., Ref.\cite{CBF})
\beq
\label{cbasis}
| n \rangle =\frac{F | n_{FG} \rangle}{\langle n_{FG}| F^{\dagger} F | n_{FG} \rangle^{1/2}} \ .
\eeq    
The operator $F$, embodying the correlation structure induced by the NN
interaction, is written in the form
\begin{equation}
F=\mathcal{S}\prod_{ij} f_{ij} \  ,
\end{equation}
where $\mathcal{S}$ is the symmetrization operator accounting for the fact that, 
as the operator structure of the two-body correlation functions $f_{ij}$ 
reflects the complexity of the NN potential,  in general
$[f_{ij},f_{ik}] \neq 0$.

Within the CBF approach the new basis defined by Eq.(\ref{cbasis})  
is employed to perform perturbative calculations with the {\em bare} NN potential. However, the same formalism can be also exploited to obtain 
an {\em effective interaction}, suitable for use with the 
Fermi gas basis \cite{shannon,BV}. The CBF effective interaction, $v{_{\rm eff}}$, is defined by the relation
\beq
\label{def:veff}
\frac{\langle 0 | H | 0 \rangle}{\langle 0 | 0 \rangle} =
\langle 0_{FG} | K + v_{{\rm eff}}| 0_{FG} \rangle  \ ,
\eeq
where $| 0_{FG} \rangle$ and $| 0 \rangle$ denote the Fermi gas and correlated ground state, respectively, 
$H$ is the nuclear hamiltonian and $K$ is the kinetic energy operator. 

In Ref. \cite{BV}, $v_{\rm eff}$ has been derived starting from  
a truncated version of the state-of-the-art NN potential referred to as Argonne $v_{18}$ \cite{av18,av8p}. 
The effects of three- and many-nucleon interactions, which are known to play a critical role in determining 
both the spectra of few-nucleon systems \cite{PW} and the saturation properties of isospin symmetric nuclear matter \cite{akmal}, 
have been also included, following the approach originally proposed in Ref. \cite{TNI}.

The energy per particle of both symmetric nuclear matter and pure neutron matter at 
temperature $T=0$, obtained from $v_{\rm eff}$ in the Hartree-Fock approximation
\be
\label{HF:E}
\frac{E}{A} & = & K_0  +  \frac{1}{2} \int \frac{d^3p_i}{(2 \pi)^3}\frac{d^3p_j}{(2 \pi)^3}  n({\bf p}_i)n({\bf p}_j)  \\
\nonumber
 & & \ \ \ \ \ \ \ \ \ \ \ \ \ \ \ \ \ \ \ \ \ \ \ \ \ \  \times \left[ \hat{v}(0)- \hat{v}({\bf p}_i-{\bf p}_j) \right] \ ,
\ee
turns out to be in good agreement with the results of highly advanced many-body approaches \cite{BV}.

In Eq.\eqref{HF:E}, $K_0$ is the energy of the non interacting system, $n({\bf p})$ is the Fermi distribution function,
\beq
\label{HF2}
\hat{v}(0) = \frac{1}{\rho}  \int d^3x  \ \sum_{\lambda_i,\lambda_j} \langle \lambda_i \lambda_j | v_{\rm eff} | \lambda_i \lambda_j \rangle \ ,
\eeq   
and
\beq
\label{HF3}
\hat{v}({\bf p}) = \frac{1}{\rho}  \int d^3x \ {\rm e}^{i {\bf p} \cdot {\bf x}} \sum_{\lambda_i, \lambda_j}  \langle \lambda_i \lambda_j | v_{\rm eff} | \lambda_j \lambda_i \rangle \ ,
\eeq   
where ${\bf x}$ is the distance between the two interacting nucleons and $\rho=A/V$, $V$ being the normalization volume, is the density. The label $\lambda_i$ 
denotes the spin-isospin quantum numbers  specifying the state of the $i$-th particle.   

\subsection{Landau parameters}
\label{landau_par}
The theory of normal Fermi liquid developed by Landau in the 1950s \cite{L1,L2} describes a uniform system of interacting fermions near $T=0$. 

The basic tenet of Landau's approach is that, as interactions are adiabatically switched on, the states of the non interacting system smoothly evolve into 
interacting states. As a consequence, the low energy excitations of the liquid, dubbed {\em quasiparticles}, retain some of the essential properties of the  
non interacting Fermi gas. 

The distribution of quasiparticles with momentum ${\bf p}$ and spin ${\bm \sigma}$ (from now on, the isospin 
degree of freedom will be dropped, as our discussion will focus on pure neutron mater)
is described by the Fermi distribution
\begin{equation}
\label{L1}
n_{\bm p\bm\sigma}= [ 1+ {\rm e}^{\beta(\epsilon_{\bm p\bm \sigma}-\mu) }]^{-1} \ ,
\end{equation}
where $\mu$ is the chemical potential and $\beta = 1/k_BT$, $k_B$ being the Boltzmann constant.

The excitation of the system is measured by the departure $\delta n_{\bm p\bm\sigma}$ from the ground state
(i.e. $T=0$) distribution
\begin{equation}
\delta n_{\bm p\bm \sigma}= n_{\bm p\bm \sigma}- n_{\bm p\bm \sigma}^0 \ . 
\end{equation}
Note that Eq.(\ref{L1}) is an implicit equation for $n_{\bm p\bm\sigma}$, as the quasiparticle energy $\epsilon_{\bm p\bm \sigma}$ depends on the distribution through
\begin{equation}
\epsilon_{\bm p\bm \sigma}=\epsilon_{\bm p\bm \sigma}^0 + \sum_{\bm p'\bm \sigma'}f_{\bm \sigma\bm \sigma'\bm p \bm p^\prime}\delta n_{\bm p^\prime \bm \sigma^\prime} \ .
\end{equation}
In the above equation, $f_{\bm \sigma\bm \sigma'\bm p\bm p'}$ describes the interaction between two quasiparticles with momentum and spin $\bm p \bm\sigma$ and $\bm p'\bm\sigma'$, 
 respectively, while near the Fermi surface $\epsilon_{\bm p\bm \sigma}^0$ can be written as
\begin{equation}
\epsilon_{\bm p\bm \sigma}^0= \mu + v_F(p -p_F) \ ,
\end{equation}
where the Fermi velocity is defined as
\beq
v_F = \frac{p_F}{m^\star} = \left( \frac{\partial \epsilon_{\bm p\bm \sigma}^0}{\partial p} \right)_{p=p_F}  \ ,
\eeq
and $m^\star$ and $p_F$ are the quasiparticle effective mass and the Fermi momentum, respectively. 

The quantity $f_{\bm \sigma\bm \sigma'\bm p\bm p'}$ incorporates both the spin-dependence and the non-central 
nature of the NN interaction in the neutron-neutron channel. As a consequence, it can be cast in the form \cite{DH}
\begin{equation} \label{L2}
f_{\bm \sigma\bm \sigma'\bm p\bm p'}= f_{\bm p\bm p'}+g_{\bm p\bm p'}\bm\sigma\cdot\bm\sigma' + h_{\bm p\bm p'}S_{12}({\bf q}) \ ,
\end{equation}
where $\bm q=\bm p-\bm p'$ and 
\begin{equation}
S_{12}({\bf q})=3 \frac{ (\bm \sigma \cdot {\bf q} )(\bm \sigma^\prime \cdot {\bf q}) }{ {|{\bf q}|^2} }-(\bm \sigma\cdot \bm\sigma') \ .
\end{equation}
The role of additional non-central contributions has been analyzed in Ref. \cite{SF}. However, their inclusion is not expected to  
significantly affect the results discussed in this paper. 

Within the approximations underlying Landau's theory, all momenta are restricted to the Fermi surface, where the quasiparticle concept is well defined. 
Hence,  one can set $|{\bf p}|=|{\bf p^\prime}|=p_F$, and describe the dependence on 
$\cos \xi = \bm p \cdot \bm p'/p_F^2$ by expanding $f_{{\bm p\bm p'}}$,  $g_{\bm p\bm p'}$ and $h_{\bm p\bm p'}$ in series of Legendre polynomials.
For example, the expansion of  $f_{{\bm p\bm p'}}$ reads
\begin{equation}
f_{\bm p\bm p'}=\sum_{\ell=0}^{\infty}f_\ell P_\ell(\cos\xi) \ .
\end{equation}

Up to quadratic terms in $\delta n_{\bm p\bm\sigma}$, the excitation energy of the system is given by the expression 
\begin{equation}
\delta E= \sum_{\bm p\bm\sigma} \epsilon_{\bm p\bm\sigma}^o \delta n_{\bm p\bm\sigma} + \frac{1}{2}\sum_{\bm p\bm\sigma \bm p'\bm\sigma'}f_{\bm\sigma \bm\sigma'\bm p\bm p'}\delta n_{\bm p\bm\sigma}\delta n_{\bm p'\bm\sigma'}  \ ,
\end{equation}
showing that the quasiparticle interaction can be identified with the the second functional derivative of the total energy with respect to $\delta n_{\bm p\bm\sigma}$
\begin{equation}
f_{\bm \sigma\bm\sigma'\bm p \bm p'}=\frac{\delta^2E}{\delta n_{\bm p\bm \sigma}\delta n_{\bm p'\bm \sigma'}} \ .
\end{equation}

The above equation can be exploited to obtain $f_{\bm \sigma\bm\sigma'\bm p \bm p'}$ from the energy computed in Hartree-Fock approximation with 
the CBF effective interaction [see Eq.(\ref{HF:E})]. Combining the resulting expression,  
\begin{equation}
f_{\bm \sigma\bm\sigma'\bm p \bm p'}  = \left[ \hat{v}(0)- \hat{v}(\bm p-\bm p') \right]  , 
\end{equation}
with Eqs. (\ref{HF2}), (\ref{HF3}) and (\ref{L2}), one can then obtain the coefficients $f_\ell$,  $g_\ell$ and $h_\ell$ appearing in the expansions of 
$f_{{\bm p\bm p'}}$,  $g_{\bm p\bm p'}$ and $h_{\bm p\bm p'}$.

The Landau parameters $F$, $G$ and $H$ are dimensionless quantities obtained multiplying $f_{{\bm p\bm p'}}$,  $g_{\bm p\bm p'}$ and $h_{\bm p\bm p'}$  by 
the density of states at the Fermi surface, $N_0= m^\star p_F/\pi^2$.

\begin{figure}[t]
\begin{center}
\includegraphics[scale=0.55]{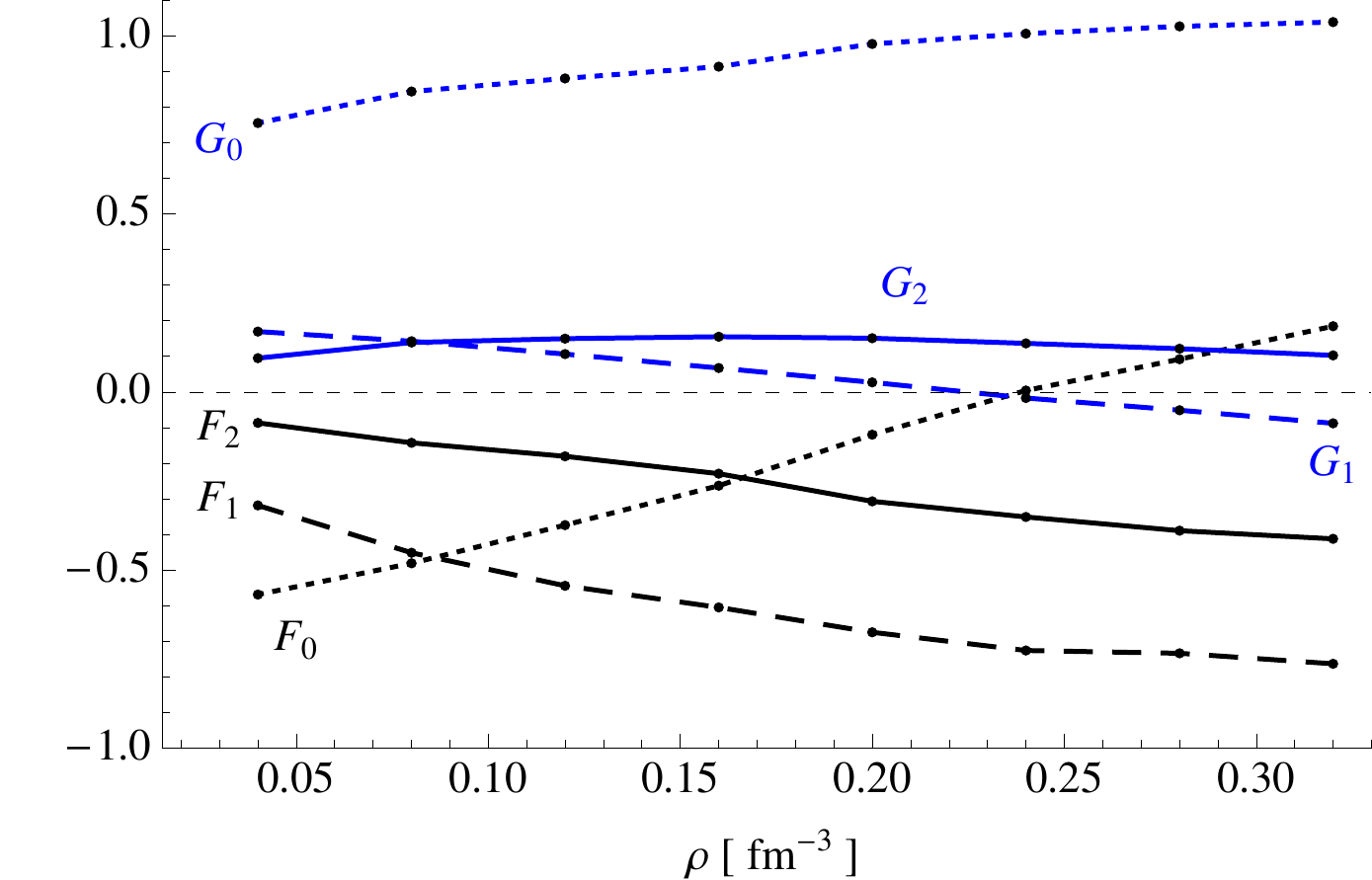}
\caption{Density-dependence of the Landau parameters $F_\ell$ and $G_\ell$ of pure neutron matter obtained from the matrix elements of the 
CBF effective interaction.}\label{LF1}
\end{center}
\end{figure}

\begin{figure}[t]
\begin{center}
\includegraphics[scale=0.55]{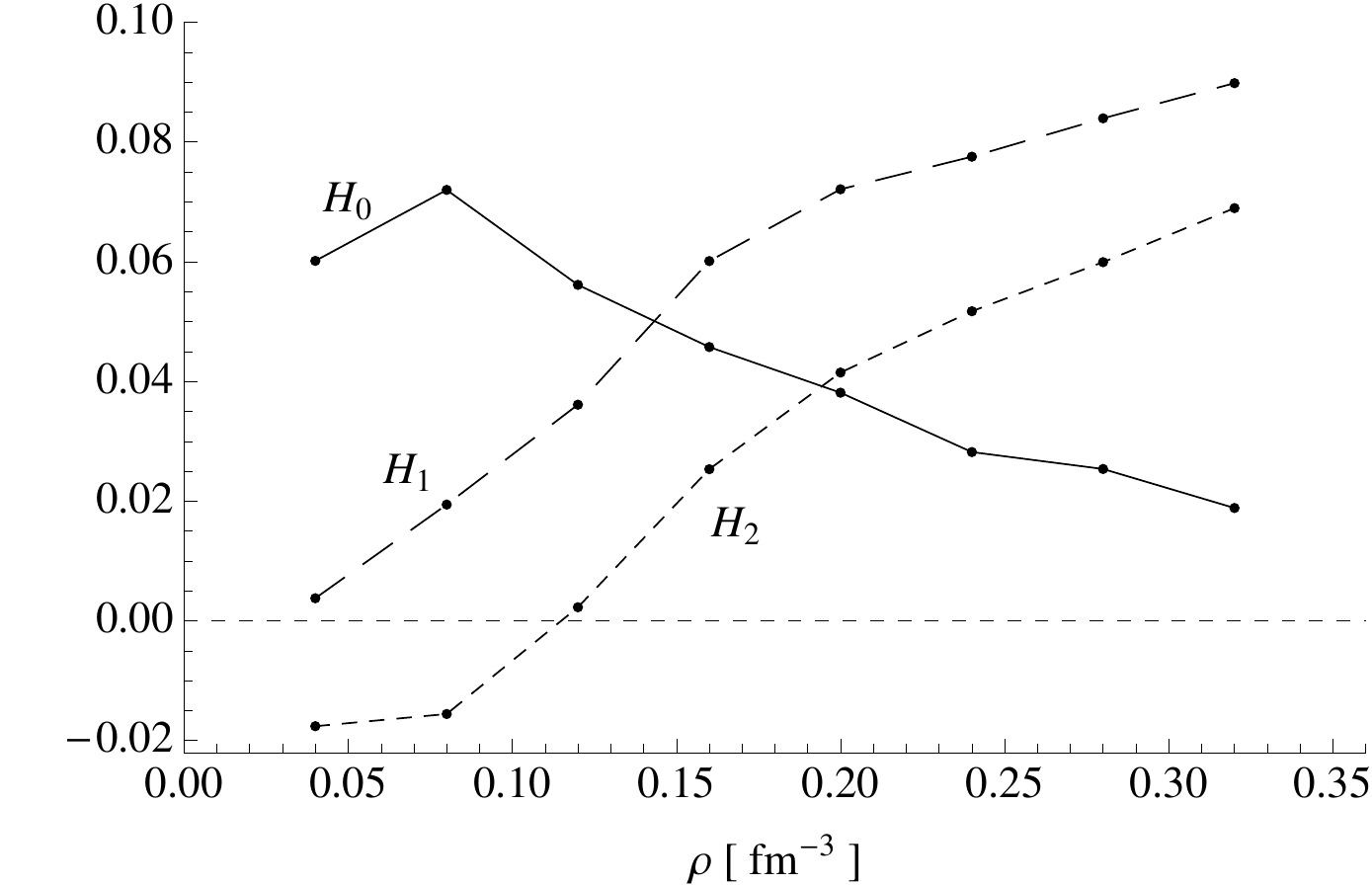}
\caption{Same as in Fig.  \ref{LF1}, but for the Landau parameters $H_\ell$.}\label{LF2}
\end{center}
\end{figure}

The values of $F_\ell$, $G_\ell$ and $H_\ell$ obtained from the matrix elements of the CBF effective interaction are listed in Table \ref{LT1} and 
\ref{LT2}, for $\ell=0,1,2$. Their density-dependence is displayed in Figs. \ref{LF1} and \ref{LF2}. 

\begin{table}[h!]
\begin{center}
\begin{tabular}{ c c c c c c c  }
\hline
\hline
 $\rho \  [{\rm fm}^{-3}]$                    & $F_0$ &  $F_1$  & $F_2$ &  $G_0$  &  $G_1$ & $G_2$\\
\hline
0.04 & -0.569 & -0.319 &-0.087      & 0.755 & 0.169& 0.094\\
0.08 &-0.481 & -0.452 & -0.143      & 0.844 & 0.142& 0.138\\
0.12 &-0.374 & -0.544  & -0.180     & 0.881 &  0.107&0.150 \\
0.16 &-0.263 & -0.605 & -0.230  & 0.914 & 0.067&0.155 \\
0.20 &-0.120 & -0.675 &  -0.307  & 0.978 & 0.027& 0.150\\
0.24 &0.004 & -0.726  & -0.351   & 1.006 & -0.017&0.136 \\
0.28 &0.092 & -0.734  &  -0.390  & 1.027 & -0.052&0.121 \\
0.32 &0.184 & -0.763  &  -0.412    & 1.039 & -0.088&0.102 \\
\hline
\hline
\end{tabular}
\caption{Landau parameters $F_\ell$ and $G_\ell$ of pure neutron matter obtained from the matrix elements of the 
CBF effective interaction.
}\label{LT1}
\end{center}
\end{table} 
\begin{table}[h!]
\begin{center}
\begin{tabular}{c c c c}
\hline
\hline
$\rho \ [{\rm fm}^{-3}]$ &  $H_0$ & $H_1$ & $H_2$ \\
\hline
 0.04 & 0.060 & 0.004 & -0.018\\
 0.08 & 0.072 & 0.019 & -0.016\\
 0.12 & 0.056 & 0.036 & 0.002\\
 0.16 & 0.046 & 0.060 & 0.025\\
 0.20 & 0.038 & 0.072 & 0.041\\
 0.24 & 0.028 & 0.078 & 0.052\\
 0.28 & 0.025 & 0.084 & 0.060\\
 0.32 & 0.019 & 0.090 & 0.069\\
\hline
\hline
\end{tabular}
\caption{Same as in Table \ref{LT1}, but for the Landau parameters $H_\ell$. }\label{LT2}
\end{center}
\end{table}

\section{Equilibrium properties of pure neutron matter}
\label{static}

Before discussing the application of Landau theory to the calculation of the dynamic structure functions, in this Section we report the results of the 
calculations of a variety of equilibrium properties of neutron matter at $T=0$. 

We will focus on  isothermal compressibility, $\mathcal{\chi}^\rho$, effective mass, $m^*$, and magnetic susceptibility, $\chi^\sigma$, which, in the static limit, can be related to the density, 
energy-density and spin-density responses, respectively. As these quantities can be obtained both  from matrix elements of the CBF effective interaction and from the Landau parameters 
listed in Tables \ref{LT1} and \ref{LT2}, the analysis discussed in this Section provides a valuable consistency test of our approach. 

Compressibility and effective mass can be computed from the expressions
\begin{eqnarray}
\label{cbf:K}
& & \mathcal{\chi}^\rho =  - \frac{1}{V} \ \left(\frac{\partial V}{\partial P}\right) \ ,\\
& &  m^{*} =  \left( \frac{1}{p}\frac{d e}{d p} \right)^{-1}_{p=p_F} \ ,
\end{eqnarray}
where the pressure is defined as $P = -\partial E / \partial V$, with $E$ given by Eq.(\ref{HF:E}), and $e(p)$ is the single particle spectrum, that can be consistently obtained from the 
CBF effective interaction within the Hartree-Fock approximation \cite{Benhar_Farina}. 

The corresponding expressions in terms of Landau parameters are
\begin{eqnarray}
\label{lan:K}
& & \mathcal{\chi}^\rho = \frac{1}{\rho^2}\frac{N_0}{1+F_{0}} \ , \\
& & m^*=m\left(1+\frac{F_1}{3}\right) \ .
\end{eqnarray}

The magnetic susceptibility can also be computed using the CBF effective interaction, following the procedure described in Ref. \cite{pfarr}. 

The energy of 
matter with spin-up (spin-down) neutron densitiy $\rho_\uparrow$ ($\rho_\downarrow$), that can be obtained through a straightforward generalization of
Eq.(\ref{HF:E}), can be cast in the form \cite{shannon_2}
\beq
E(\rho, \alpha)=E_0(\rho)+E_{\sigma}(\rho)\alpha^{2} \ ,
\eeq
with $\alpha = (\rho_\uparrow$ - $\rho_\downarrow)/\rho$. 

In the presence of a uniform magnetic field ${\bf B}$,  the above equation becomes 
\beq
E_B(\rho,\alpha) = E(\rho,\alpha) - \alpha \mu_0 B \ ,
\eeq
where $B$ denotes the magnitude of the external field, the direction of which is chosen as spin quantization axis, and $\mu_0$ is
the magnetic moment of a neutron in free space. From the definition of total magnetization, $M =  \mu(\rho_\uparrow$ - $\rho_\downarrow) = \chi^\sigma B$, 
 it follows that, at equilibrium,  
\beq
\chi^\sigma = \mu_0^2\left(\frac{\partial^2 E}{\partial \alpha^2}\right)^{-1}_{\alpha=0}\rho \ .
\eeq

In the case of spherically symmetric interactions, the expression of the magnetic susceptibility obtained from Landau's theory reads
\beq
\label{chi:L1}
\chi^\sigma_0=\mu_0^2\frac{1}{1+G_0} \ .
\eeq
When non-central forces are present, the neutron magnetic moment is a tensor of the form 
\beq
\mu_{ij} = {\tilde \mu}_0 \delta_{ij} + \frac{3}{2} \mu_T \left( \frac{p_i p_j}{|{\bf p}|^2} - \frac{\delta_{ij}}{3} \right) \ ,  
\eeq
where ${\tilde \mu}_0$ is the medium modified neutron magnetic moment.
 Neglecting the contributions involving $\mu_T$, the expression of the susceptibility in terms of Landau's 
parameters is \cite{Haens_2,OHP}
\beq
\label{chi:L2}
\chi^\sigma={\tilde \mu}_0^2\frac{1}{1+G_0}\Big[1+\frac{1}{8}\frac{1}{1+G_0}\frac{(H_0-H_1)^2}{1+G_2/5} \Big] \ ,
\eeq
The effect of the tensor interaction turns out to be negligibly small. Using the values of $G_\ell$ and $H_\ell$ reported in Tables \ref{LT1} and \ref{LT2}, and
setting ${\tilde \mu}_0 = \mu_0$, one finds that $\chi^\sigma$ given by the above equation differs from $\chi^\sigma_0$ of Eq.(\ref{chi:L1})
by $\sim 0.01\%$. 
\begin{figure*}
\begin{center}
\includegraphics[scale=0.5]{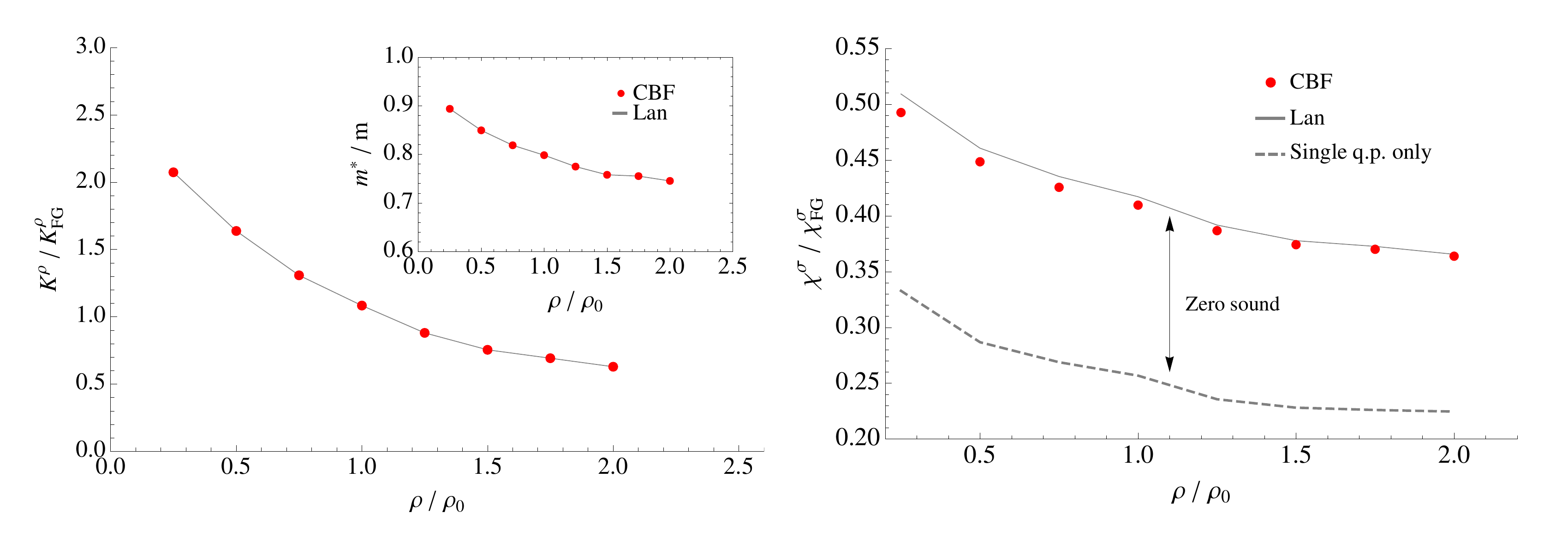}
\vspace*{-0.3 cm}  
\caption{Left panel: Compressibility of neutron matter, normalized to its Fermi gas value, as function of density in units of $\rho_0=0.16 $ fm$^{-3}$. 
The dots and the solid line correspond to the results obtained from Landau's theory [Eq.(\ref{lan:K})] and the equation of state computed using the CBF effective interaction [Eq.(\ref{cbf:K})], respectively. Right panel: same as in the left panel, but for for the spin susceptibility, $\chi^\sigma$. The dashed line shows the 
susceptibility obtained from the dynamic spin structure function including only the incoherent contribution.  The inset of the left panel shows the density dependence of the ratio between effective 
and bare neutron mass.}   
\label{fig:static}      
\end{center}
\end{figure*}

In Fig. \ref{fig:static} the neutron matter compressibility, effective mass and magnetic susceptibility obtained from Landau theory (dots) and from the CBF 
effective interaction (solid lines) are compared as a function of density. Note that $\mathcal{\chi}^\rho$ and $\chi^\sigma$ are normalized to their Fermi gas values, while the effective mass is given in units of the bare neutron mass.

It appears that the results obtained from the two schemes are in close agreement with one another over the whole range of densities, extending to twice the nuclear 
matter equilibrium density, $\rho_0 = 0.16 \ {\rm fm}^{-3}$. 

The left panel shows that in the low-density region, where the effective interaction is predominantly attractive, the compressibility is larger than in the non interacting Fermi gas. 
Increasing $\rho$, the effective interaction changes sign and the picture is reversed.

On the other hand, the right panel shows that the inclusion of interaction effects leads to a suppression of the spin susceptibility, with respect to the Fermi gas value, at all densities. 
This behavior suggests that repulsive interactions are dominant in spin-density channel. 

\begin{figure*}
\begin{center}
\includegraphics[scale=0.6]{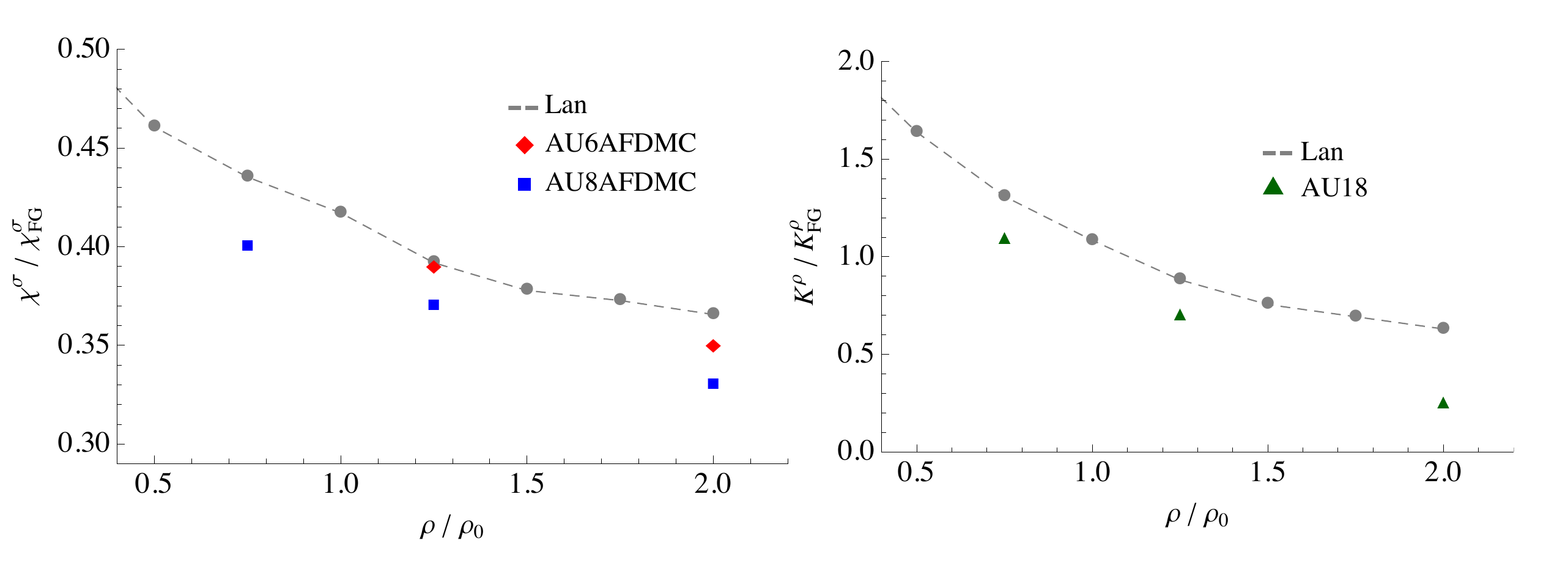}
\vspace*{-0.5 cm}       
\caption{Left panel: comparison  between the spin susceptibility computed within Landau's theory and the corresponding 
results obtained in Ref. \cite{Fantoni} using the Auxiliary Field Diffusion Monte Carlo approach. Right panel: comparison  between the compressibility
 computed within Landau's theory and the corresponding results obtained in Ref. \cite{akmal} using the variational FHNC-SOC approach.}   
\label{fig:static2}      
\end{center}
\end{figure*}

In the left panel of Fig. \ref{fig:static2}, the magnetic susceptibility of neutron matter computed within Landau theory is compared to the results
 of Ref. \cite{Fantoni}, obtained using the Auxiliary Field Diffusion Monte Carlo approach and nuclear hamiltonians including the truncated $v_6^\prime$ and $v_8^\prime$ \cite{av8p} forms of 
the Argonne $v_{18}$ potential \cite{av18}, supplemented with the Urbana IX three-nucleon potential \cite{UIX}. In the right panel, we compare the density dependence of
the compressibility resulting from our calculations to the results of the variational calculations of Ref. \cite{akmal},  carried out using the full Argonne  $v_{18}$ NN potential 
and the Urbana IX three-nucleon potential. Our results appear to be in reasonable agreement with those obtained from highly refined theoretical approaches, 
the differences at large density being likely to be ascribable to the different treatment of three-nucleon forces, which are known to play a
critical role at $\rho > \rho_0$. 
      
\section{Response functions at zero temperature}
\label{energy}
In the long wavelength limit, corresponding to $|{\bf q}| \rightarrow 0$, where ${\bf q}$ is the momentum transfer, the dynamic response is determined 
by the physics of the Fermi surface. 

At low $|{\bf q}|$, the relevant microscopic degrees of freedom, i.e. the particle-hole excitations, can be treated in a ring-like 
approximation scheme using the formalism of Landau theory \cite{IW}. The effect of long range correlations, which are known to be dominant in this region, can be 
included in a fully consistent fashion, taking into account the incoherent and coherent contributions to the response on equal footing. 

In this Section, we will
outline the application of Landau theory to the calculation of the weak dynamic structure function, 
i.e. the correlation function of the hadronic weak current, needed to obtain the neutrino mean free path in neutron matter to be discussed in Section \ref{mfp}.  

For a generic operator $O_{{\bf q}}$, the correlation function can be written as the sum of  
two contributions, according to 
\begin{eqnarray}
\label{corfun}
S({\bf q},\omega) & = & \int \frac{dt}{2\pi} \ \langle 0 | O^{\dagger}_{{\bf q}}(t) O^{\dagger}_{{\bf q}}(0)  | 0 \rangle \\ 
\nonumber
& = & \sum_{ph} \langle 0|O^{\dagger}_{{\bf q}}|ph\rangle\langle ph|O_{{\bf q}}|0\rangle \delta (\omega-\omega_{n0})  + S_{{\rm MP}} \ ,
\end{eqnarray}
where $\omega_{n0} = E_n - E_0$, $E_0$ and $E_n$ being the energies of the states $|0\rangle$ and $|n\rangle$, respectively. The sum in the first term includes all one 
particle-one hole eigenstates of the hamiltonian,  while the contribution of more complex {\em multipair} states 
is included in $S_{{\rm MP}}$ \cite{MP}. 

The linear response function can be written in the same form
\be
\nonumber
\label{chifun}
\chi({\bf q},\omega)& = & \sum_{ph} \langle 0|O^{\dagger}_{{\bf q}}|ph\rangle\langle ph|O_{{\bf q}}|0\rangle \\ 
 & & \ \ \ \ \ \ \times \frac{2 \omega_{n0}}{(\omega+i\eta)^2-\omega_{n0}^2} +  \chi_{{\rm MP}} \ ,
\ee
with $\eta = 0^+$. The link between Eqs.(\ref{corfun}) and (\ref{chifun}) is provided by the fluctuation-dissipation theorem, stating that, at $T=0$,
\beq
\label{FDT}
S({\bf q},\omega) =-\frac{1}{\pi} \textrm{Im} \ \chi({\bf q},\omega). 
\eeq
The theoretical calculation of the first term of Eq.(\ref{chifun}) is based on the linearized Landau-Boltzmann transport equation in the presence of 
an external probe transferring low momentum. Working within this framework, one can obtain the explicit expression of the response to a scalar probe, 
$\chi^{\rho \rho}({\bf q},\omega)$, involving the two Landau parameters $F_0$ and $F_1$ \cite{IW}
\beq
\chi^{\rho \rho}({\bf q},\omega)=\frac{N_0}{V}\frac{\Omega(\lambda)}{1+[F_0+\lambda^2\frac{F_1}{1+F_1/3}]\Omega(\lambda)} \ .
\label{response}
\eeq
In the above equation 
\beq
\label{def:lambda}
\lambda= \frac{\omega}{v_F|{\bf q}|} ,
\eeq
is the speed of the perturbation in units of the Fermi velocity, while $\Omega(\lambda)$ is the Lindhardt (or polarization) function of the free Fermi gas at $T=0$ \cite{FW}. The extension 
to include more than two Landau parameters is straightforward \cite{thesis}. 

It clearly appears that all interaction effects are described by the ``potential'' in square bracket, that can be written in terms of Landau parameters and is therefore constrained by the static properties of matter. Note that for $\omega/|{\bf q}| <v_F$ the denominator of Eq.(\ref{response}) can vanish, thus indicating the presence of a resonance between the external perturbation and the single quasi particle excitation. An undamped mode can exist in the region corresponding to $\lambda>1$, where the imaginary part of the response reduces to a delta function.

\subsection{Density and spin responses}

\begin{figure*}
\begin{center}
\includegraphics[scale=0.6]{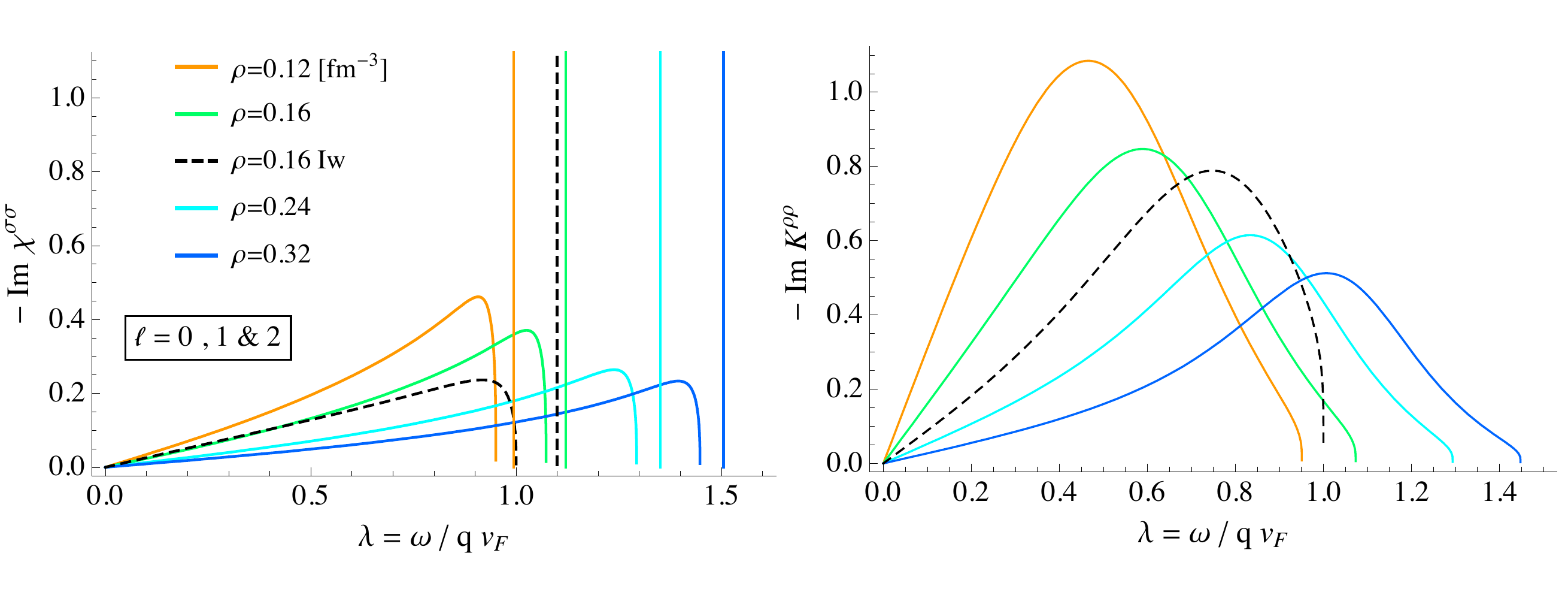}
\vspace*{-.4 cm}       
\caption{Left panel: spin-density structure function of neutron matter, as function of $\lambda$, defined in Eq.(\ref{def:lambda}), at different densities. 
The calculations have been carried out including the Landau parameters with $\ell =$ 0, 1 and 2.  
For comparison, the thick dashed line shows the results of Ref. \cite{IW}, corresponding to $\rho = \rho_0 = 0.16 \ {\rm fm}^{-3}$. Right panel: same as in the left panel, but for the density structure function.}   
\label{fig_energy_spin}      
\end{center}
\end{figure*} 
\begin{figure*}
\begin{center}
\includegraphics[scale=0.6]{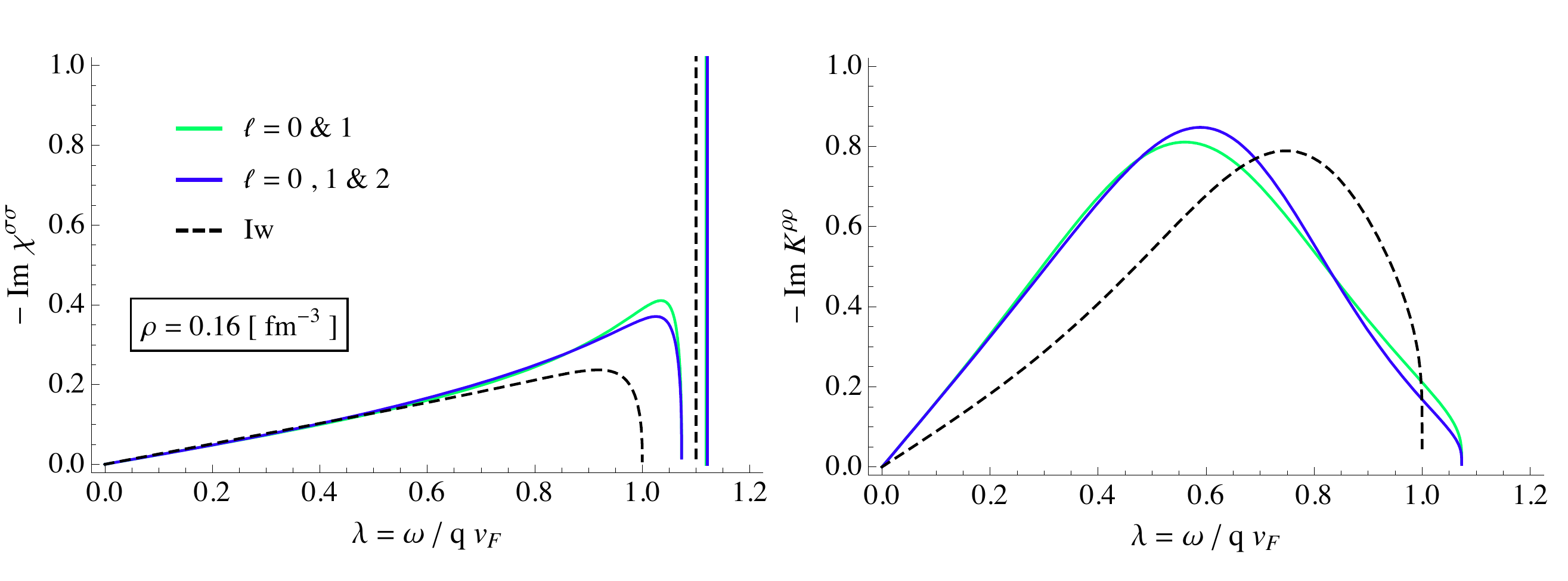}
\vspace*{-0.5 cm}       
\caption{Left panel: spin-density structure function of neutron matter at $\rho = \rho_0 = 0.16 \ {\rm fm}^{-3}$, as function of $\lambda$. The curves have been 
obtained including only the Landau parameters with $\ell =0, \ 1$ or taking into account the contribution associated with $\ell =2$.
 For comparison, the thick dashed line shows the results of Ref. \cite{IW}. Right panel: same as in the left panel, but for the density structure function.}   
\label{fig_energy_spin_2}      
\end{center}
\end{figure*} 

Neutrino-neutron scattering via the neutral weak current can be described within Weinberg-Salam theory in the low-energy limit. 
The starting point is the interaction lagrangian density 
\begin{equation}\label{L3}
{\mathcal L}_{I}(x)=\frac{G_F}{\sqrt{2}} \ \ell_\mu(x)j_Z^{\mu }(x) , 
\end{equation}
where $G_F\simeq 4.55\times 10^{-7} \  {\rm fm}^2 $ is the Fermi coupling constant, while 
\begin{equation}
\label{L8}
\ell_\mu =\overline{\psi}_{\nu} \gamma _\mu (1-\gamma _5)\psi _{\nu } \ ,
\end{equation}
and
\begin{equation}
j^\mu_Z=\frac{1}{2}\overline{\psi}_n\gamma^{\mu}(1 - C_A\gamma ^5)\psi_n \ ,
\end{equation}
are the neutrino current and the third component of the isospin current, with $C_A\simeq 1.25$, respectively.
Treating the neutrons as non relativistic particles, one can approximate the components of $j^{\mu}_Z$ according to
\begin{equation}
\label{L4}
\overline{\psi}_n{\gamma }^{\mu }\psi_n\rightarrow\psi_n^\dagger \psi_n\delta^\mu_o ,
\end{equation}
\begin{equation}\label{L5}
 \overline{\psi}_n{\gamma }^{\mu }{\gamma }^5\psi_n\rightarrow \psi_n^\dagger {\sigma }^i\psi_n\delta^\mu_i ,
\end{equation}     
where $\sigma^i$ $(i=1,2,3)$ are Pauli matrices in spin space. From Eqs. (\ref{L3})-(\ref{L5}), it follows that the time component of the neutrino current couples to the time component 
of the neutron current to give rise to density fluctuations, while the coupling between the space components of the neutrino and neutron currents leads to spin-density fluctuations.

Let  $(k_0,\bm k)$ and $(k^\prime_0,\bm k^\prime)$ be the initial and final four-momenta of the neutrino, respectively, while the corresponding neutron four-momenta 
will be denoted $(p_0,\bm p)$ and $(p^\prime_0,\bm p^\prime)$. In the non relativistic limit, the scattering rate, computed using Fermi golden rule reads
\begin{eqnarray}
\label{L6}
\nonumber
W({\bf q},\omega) & = & \frac{G_F^2 \rho}{4V} \left\{ (1+\cos\theta)\textit{\emph{S}}\left(\bm{q},\omega \right) \right. \\
\nonumber
& + &  C^2_A \left[ {\widehat{k^\prime_i}}\widehat{k}_j{+\widehat{k^\prime_j}} \widehat{k}_i+\left(1-{\cos  \theta \ }\right){\delta }_{ij}\right] S_{ij}^+(\bm q,\omega) \\
& + & \left. i C^2_A \varepsilon_{ij \ell} 
\left( \widehat{k}^\ell-\widehat{k}^{\prime \ell}\right) \ S_{ij}^- \left(\bm {q},\omega \right)  \right\} ,
\end{eqnarray}
where ${\bf q} = {\bf k} - {\bf k}^\prime$, $\omega = k_0 - k_0^\prime$, $\widehat{k}={\bf k}/|{\bf k}|$ and  $\theta$ is the neutrino scattering angle. In the above equation
 \begin{equation}
S_{ij}^\pm(\bm q,\omega)=\frac{1}{2}\left[S_{ij}(\bm q,\omega) \pm S_{ji}(\bm q,\omega) \right]  \ ,
\end{equation}
are the symmetric and antisymmetric parts of the dynamic spin structure function
\begin{equation}
\label{spin:sqw}
S_{ij}\left(\bm q,\omega \right)= \int {\frac{dt}{2\pi} \ {\rm e}^{i\omega t} \langle {\sigma }_i\left(\bm q,t\right){\sigma }_j(-\bm q,0) \rangle},
\end{equation}  
where $\langle \ldots \rangle$ denote the ground state expectation value and $\sigma_{i}(\bm k,t)$ is the Fourier transform of the spin-density operator $\psi_n^\dagger{\sigma }_i\psi_n$.

The dynamic density structure function can be written as in Eq.(\ref{spin:sqw}), replacing $\sigma_{i}(\bm q,t)$ with $\rho(\bm q,t)$, the Fourier transform of the density 
operator  $\psi_n^\dagger \psi_n$
\begin{equation}
\label{dens:sqw}
S\left(\bm q,\omega \right)= \int {\frac{dt}{2\pi} \ {\rm e}^{i\omega t} \langle \rho \left(\bm q,t\right) \rho(-\bm q,0) \rangle} \ ,
\end{equation}  
and can be obtained from the response of Eq.(\ref{response}) using the fluctuation-dissipation theorem, Eq.(\ref{FDT}).
 
Finally, in an isotropic system $S_{ij}= \delta_{ij} {\mathcal S}$, and Eq.(\ref{L6}) reduces to 
\begin{eqnarray}
\label{L10}
W({\bf q}, \omega) & = & \frac{G_F^2 \rho}{4V} \left\{ (1+\cos\theta)S\left(\bm{q},\omega \right) \right. \\
\nonumber
& & \ \ \ \ \ \ \ \ \ \ \ + \left.  C^2_A (3-\cos\theta){\mathcal S}(\bm k,\omega) \right\} \ .
\end{eqnarray}
The structure function ${\mathcal S}(\bm k,\omega)$ can be obtained from the spin response $\chi^{\sigma \sigma}$ through Eq.(\ref{FDT}).
The calculation of the spin response within Landau theory in the presence of non central interactions is discussed in Ref. \cite{Haens_1}. 
However, for our set of Landau parameters, the calculated neutrino cross section turns out to be  nearly unaffected by the inclusion of the $H_\ell$'s. 
As a consequence,  the expression of $\chi^{\sigma \sigma}$employed in this work is the same as Eq.(\ref{response}), but with the Landau 
parameters $F_\ell$ replaced by the corresponding $G_\ell$.
   
In Fig. \ref{fig_energy_spin} we report the spin (left panel) and density dynamic structure functions  of neutron matter at different densities as a function of
$\lambda$. It clearly appears that in the case of the spin structure function the single quasi particle excitations is depleted, and a zero-sound collective mode sticks out.
As shown by the difference between the solid and dashed lines of Fig \ref{fig:static}, the collective mode provides a large contribution to the 
spin susceptibility. This is a consequence of the repulsive effective interaction in the spin channel. On the other hand, in the case of the density structure function the incoherent 
contribution exhausts the compressibility sum rule \cite{pines}
\begin{equation}
\int \frac{d\omega}{\omega}  S({\bf q},\omega) = \frac{1}{2} \ \rho \chi^\rho \   ,
\end{equation}
and there is no zero-sound mode up to density $\sim 2 \rho_0$, even when $F_{0}$ change sign. 

The results displayed in Fig. \ref{fig_energy_spin} have been obtained including the Landau parameters $G_\ell$ (left panel) and $F_\ell$ (right panel) with 
$\ell =$ 0, 1 and 2.  However, as shown in Fig. \ref{fig_energy_spin_2}, the effect of adding the contributions associated with $G_2$ and $F_2$ is very  
small. 

Note that the above discussion can be readily generalized to the case of non vanishing temperature. At $T \neq 0$, the ground state expectation of 
Eqs. \eqref{spin:sqw}-\eqref{dens:sqw} are replaced by the corresponding ensemble averages.

\section{Neutrino mean free path}
\label{mfp}

The neutrino mean free path, $L_\nu$, can be obtained from the transport equation.  From the definition of the relaxation time associated with  the distribution function of
an excited state, $\tau$, 
\begin{equation}
\frac{\partial n_{\bm p\bm\sigma}}{\partial t}= - \frac{n_{\bm p\bm\sigma}}{\tau} \ ,
\end{equation}
it follows that
\begin{align}
\label{L7}
\frac{1}{L_\nu} = V \int  \frac{ d^3 k^\prime}{(2\pi)^3} & \left\lbrace W(\bm q,\omega)[1-n(\bm k^\prime)] \right. \\
\nonumber
&+ \left. W(-\bm q,-\omega)n(\bm k^\prime)\right\rbrace ,
\end{align}
where $n$ denotes the neutrino distribution function. 

Using the principle of detailed balance, stating that
\begin{equation}
W(\bm q,\omega)=e^{\beta \omega}W(\bm q,-\omega) \ ,
\end{equation}
and the definition of the scattering rate of Eq.\eqref{L10}, one can rewrite Eq.$(\ref{L7})$ in the form
\begin{align}
\label{L12}
\nonumber
\frac{1}{L_\nu} & = \frac{G_F^2}{4} \rho \int\frac{d^3 q}{(2\pi)^3} \left[ (1 +\cos\theta)S(\bm q, \omega) \right. \\
&  \ \ \ \ \ \ \ \ \ \ \ \ \  \ \ \ \  \ \ \left. +  \ C^2_A (3-\cos\theta){\mathcal S}(\bm q, \omega) \right]  \ .
\end{align}
Note that in the above equation we set $T=0$ and assumed that neutrinos be non degenerate.

\begin{figure}[h!]
\begin{center}
\includegraphics[scale=0.55]{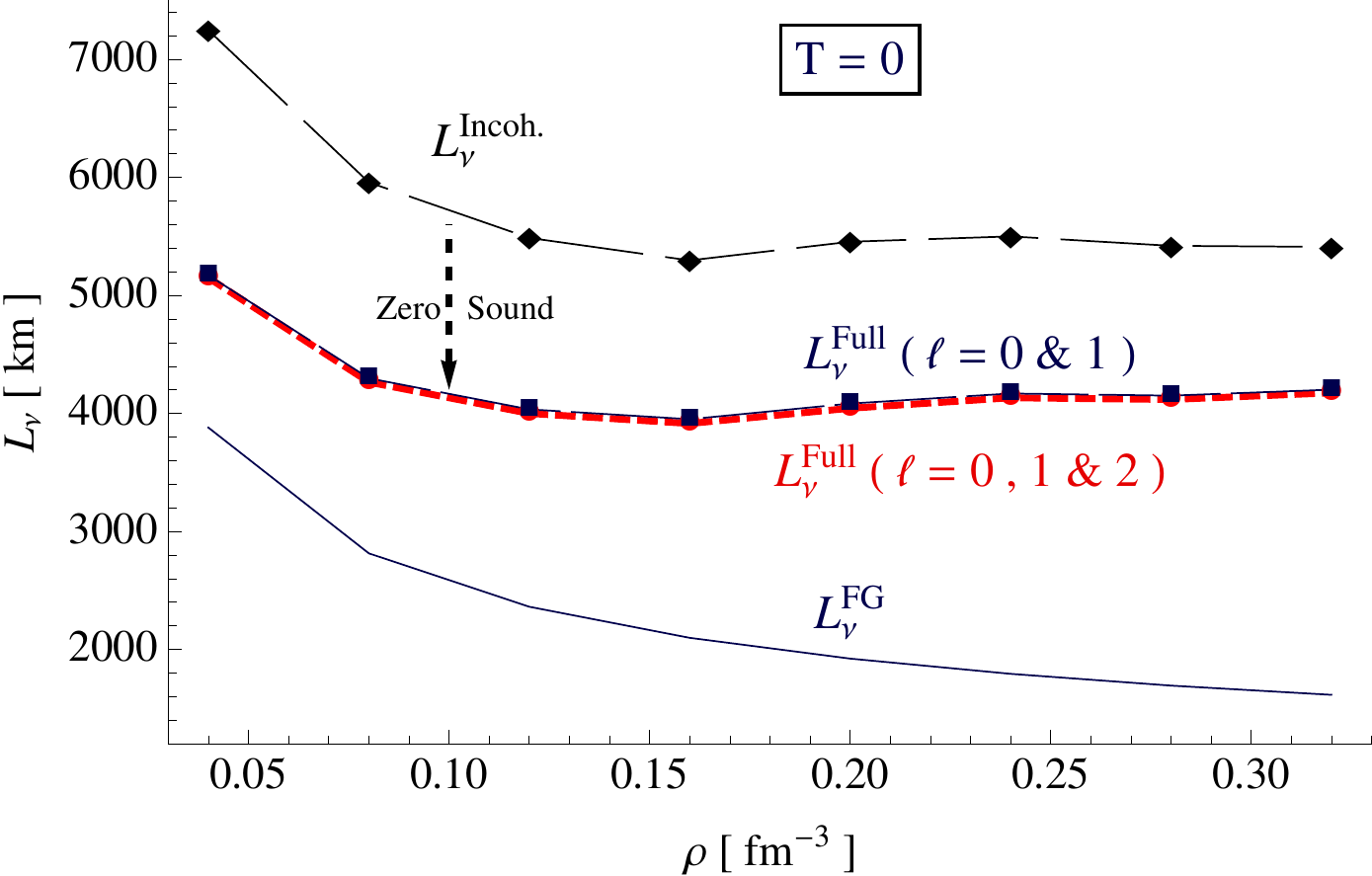}
\caption{Density dependence of the mean free path of a non degenerate neutrino with an energy $k_0 = 1$ MeV in neutron matter at $T=0$.
The density and spin-density structure functions have been computed using the Landau parameters 
$F_\ell$ and $G_\ell$ of Table \ref{LT1} with $\ell =0, \ 1$ (solid line) and $\ell =0, \ 1$ and $2$ (thick dashed line). For comparison the dashed line shows 
the mean free path in the free neutron gas.}\label{mfpvsrho_T0}
\end{center}
\end{figure}
\begin{figure}[h]
\includegraphics[scale=0.55]{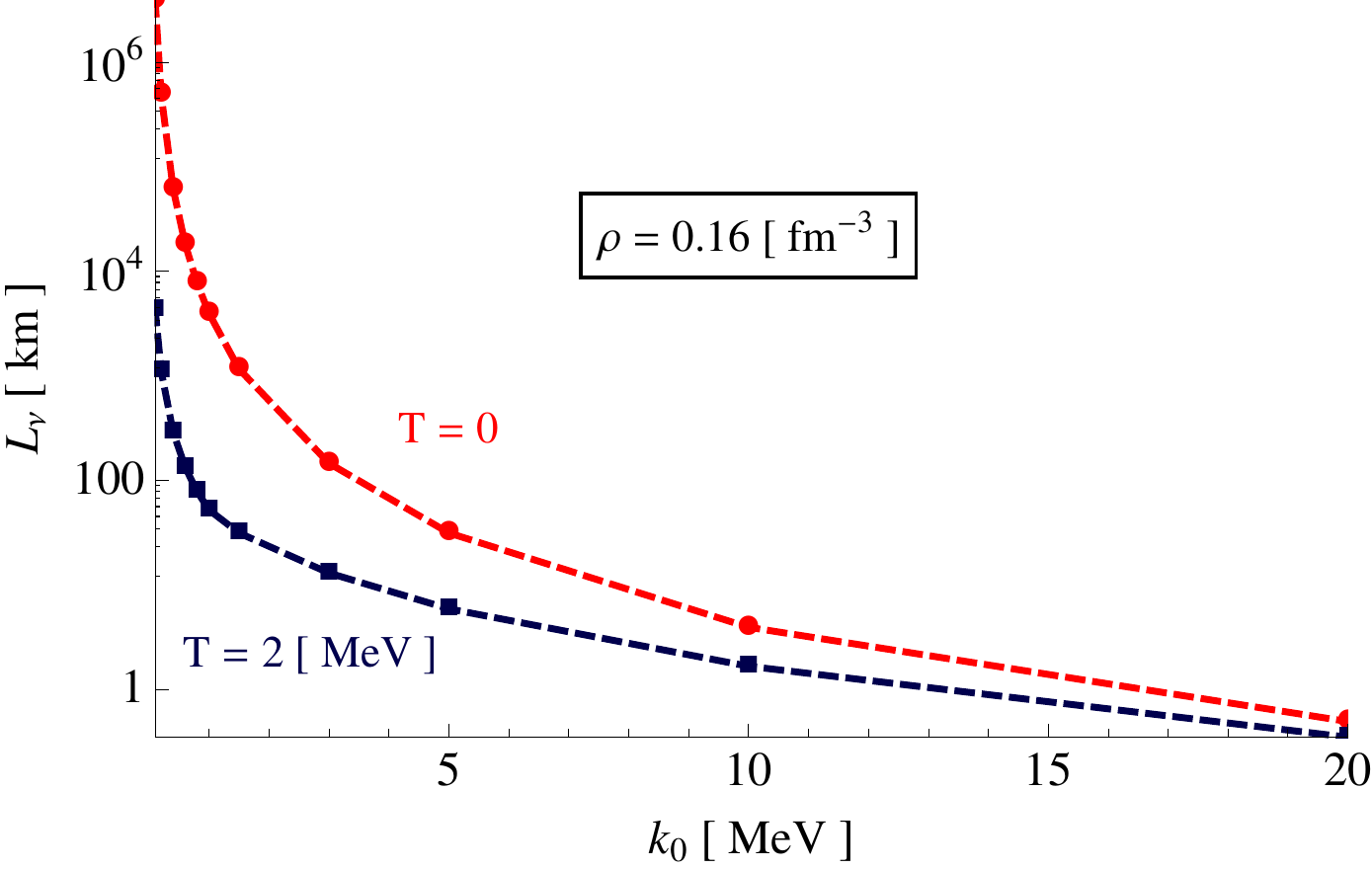}
\caption{Energy dependence of the mean free path of a non degenerate neutrino in neutron matter at different temperatures.}\label{mfpvsk0}
\end{figure}
\begin{figure*}[h]
\includegraphics[scale=0.6]{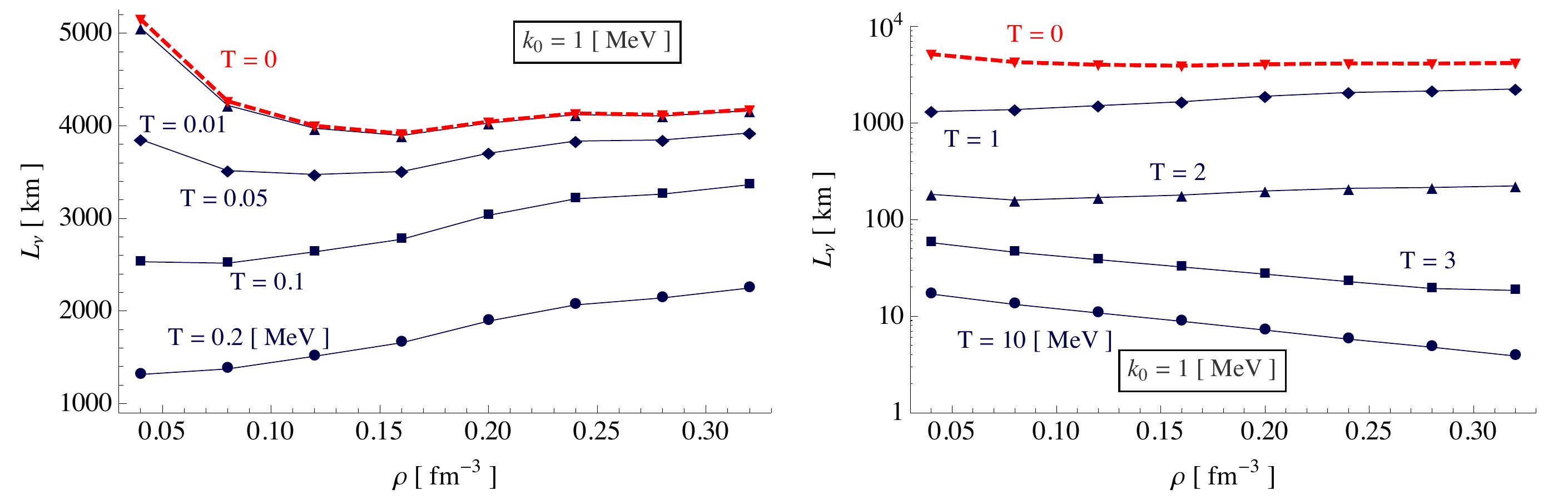}
\caption{Density dependence of the mean free path of a non degenerate neutrino with energy $k_0=1$ MeV at different temperatures.} \label{mfp_T}
\end{figure*}

Neutrino-neutron interactions can excite a one particle-one hole pair as well as a collective excitation associated with the zero-sound mode. 
Within the framework of Landau theory, the zero-sound mode propagates undamped if $\lambda \geq1$ [see Eq.\eqref{def:lambda}], i.e. if the phase velocity of the mode exceeds 
the Fermi velocity $v_F$. In order for this requirement to be fulfilled in the density (spin-density) channel, the Landau parameters $F_{0,1}$ ($G_{0,1}$) must satisfy the 
constraint \cite{hansel_condition}
\begin{equation}
F_0>0,\hspace*{.3cm}  \hspace*{.3cm} F_0 >\left| \frac{F_1}{1+F_1/3} \right| \ ,
\end{equation}
and the corresponding relations for $G_0$ and $G_1$. The results listed in Table $\ref{LT1}$ show that in the density channel the zero-sound suffers from strong Landau damping.
Hence, the spectrum of density fluctuations arises from single pair excitations only. On the other hand, this is not the case for the spectrum of spin-density fluctuations. 

The above discussion obviously implies that the inverse neutrino mean free path is the sum of two contributions, that can be obtained from Eq.\eqref{L12} singling out the 
contribution of the collective mode to the spin-density structure function. 

Figure \ref{mfpvsrho_T0} shows the density dependence of the mean free path of a non degenerate neutrino with an energy $k_0 = 1$ MeV in neutron matter at $T=0$. 
The results have been  obtained from Eq.(\ref{L12}), using the density and spin-density structure functions computed using the Landau parameters 
$F_\ell$ and $G_\ell$ listed in Table \ref{LT1} with $\ell =0, \ 1$ (solid line) and $\ell =0, \ 1$ and $2$ (thick dashed line). Comparison with the mean free path in a
free neutron gas, displayed by the dashed line, shows that inclusion of interaction effects leads to a large enhancement of $L_\nu$ over the whole density range. 

At $T\neq0$ the mean free path can still be written as in Eq.\eqref{L12}, using the appropriate expressions of the dynamic structure functions, related to the corresponding response 
functions through [compare to Eq.\eqref{FDT}]
\begin{align}
\nonumber
S({\bf q},\omega) =-\frac{1}{\pi} \frac{1}{1-{\rm e}^{-\beta \omega}} \textrm{Im} \ \chi^{\rho \rho}({\bf q},\omega) ,  \\  
{\mathcal S}({\bf q},\omega) =-\frac{1}{\pi} \frac{1}{1-{\rm e}^{-\beta \omega}} \textrm{Im} \ \chi^{\sigma \sigma}({\bf q},\omega) , 
\end{align}
where $\chi^{\rho \rho}$ and $\chi^{\sigma \sigma}$ denote the density response and the diagonal component of the spin-density response tensor, respectively.

It has to be kept in mind, however, that the description of the neutron matter response discussed in this work only applies to the regime in which 
collisions between thermally excited quasiparticles can be neglected, defined by the requirement $\omega \gg \tau^{-1}_c$, where $\tau^{-1}_c \approx T^2/T_F$, $T_F$ being the 
Fermi temperature, is the thermal collision rate. As the zero-mode frequency is $\sim T$, the collisionless regime corresponds to $T \ll T_F$.  

Figure \ref{mfpvsk0} shows the energy dependence of the mean free path in neutron matter at density $\rho=0.16$, 
corresponding to the Fermi temperatures $T_F = 35$ MeV. 
The upper  and lower curves have been obtained setting the temperature to $T=0$ and $2$ MeV, respectively. 

The dependence of the mean free path of a neutrino with energy $k_0=1$ MeV upon both temperature and 
matter density is illustrated in Fig. \ref{mfp_T}. Note that, as the the density range $0.04 \leq \rho \leq 0.32 \ {\rm fm}^{-3}$ corresponds to Fermi temperatures $14 \leq T_F  \leq 55 \ {\rm MeV}$, the collisionless condition $T \ll T_F$ is always satisfied.

\section{Conclusions}
\label{concl}

We have studied neutrino interactions in neutron matter within the framework of Landau theory of normal Fermi liquids. The values of the Landau parameters
employed in our calculations have been obtained from the matrix elements of the effective interactions recently derived in Ref. \cite{BV} using the formalism 
of correlated basis functions and the cluster expansion technique. 

Our estimates of the static properties of neutron matter turn out to be in fairly good agreement with the results of both Monte Carlo \cite{Fantoni} and
variational \cite{akmal} calculations. In the case of spin susceptibility, the difference between our results and those reported in Ref. \cite{Fantoni}, the
size of which can be used to gauge the theoretical uncertainty, is significantly smaller than the quenching arising from interaction effects.

The calculated dynamic structure functions include contributions from both quasiparticle excitations and the collective zero-sound mode. However, density 
fluctuations are strongly damped, and the $\delta$-function peak associated with the zero-sound mode is only visible in the spectrum of spin-density 
fluctuations.

The dynamic structure functions have been used to obtain the neutrino mean free path in neutron matter, which plays an important role in 
determining neutron star evolution. The calculations, performed assuming non degenerate neutrinos, have been carried out over a broad 
range of densities and extended to finite temperatures in the region $T<10$ MeV, where the collisionless approximation underlying our
approach can be safely applied. 

The results show that interaction effects give rise to a large enhancement of the mean free path, corresponding to a suppression of the 
response, providing a measure of the neutrino-neutron cross section. 

The effective interaction of Ref. \cite{BV} has been recently improved with the inclusion of three-body cluster contributions, which allows
for a more realistic treatment of three-nucleon interactions, based on microscopic potential models \cite{LLB}. A systematic comparison between 
the results obtained from the correlated Hartee-Fock and correlated Tamm-Dancoff schemes developed in Refs. \cite{shannon,Benhar_Farina,LLB}
and those obtained from Landau theory using the same dynamics will help to shed light on a number of unresolved issues, such as the role 
of multipair excitations. Additional information will also come will come from the comparison to the sum rules of the weak response computed 
using the AFDMC approach \cite{AFDMC}. 

\acknowledgments
This work was partially supported  by INFN, under grant MB31, and MIUR PRIN, under grant 
``Many-body theory of nuclear systems and implications on the physics of neutron stars''.

\end{document}